\documentclass[a4paper, 11pt]{article}
\usepackage[pdftex, a4paper]{geometry}
\usepackage{graphicx}				
				
\usepackage{fancyhdr}
\usepackage{amsmath}
\usepackage{amsfonts}
\usepackage{dsfont}
\usepackage{mathrsfs}
\usepackage{amssymb}
\usepackage{bbm}
\usepackage{epsfig}
\usepackage[english]{babel}
\usepackage[all]{xy}
\usepackage{color}

\newtheorem{theorem}{Theorem}
\newtheorem{definition}{Definition}

\newtheorem{lemma}{Lemma}
\newtheorem{remark}{Remark}
\newtheorem{proposition}{Proposition}

\title{\bf  Network Synchronization with Convexity\thanks{This work has been supported in part
by the Knut and Alice Wallenberg Foundation, the Swedish Research
Council, and KTH SRA TNG.    A preliminary version of this work has been presented at The 51st Annual Allerton Conference on Communication, Control, and Computing, Oct. 2013 \cite{allerton}.}}
\date{}

\author{Guodong Shi, Alexandre Proutiere,  and  Karl Henrik Johansson\thanks{G. Shi is with the Research School of Engineering,  College of Engineering and Computer Science, The Australian National University, Canberra ACT 0200, Australia. A. Proutiere and K. H. Johansson   are with the ACCESS Linnaeus Centre, School of Electrical Engineering,
KTH Royal Institute of Technology, Stockholm 10044, Sweden.
       Email: {\small guodong.shi@anu.edu.au, alepro@kth.se, kallej@ee.kth.se}}
}

\begin{document}

\maketitle

\begin{abstract}
In this paper, we establish  a few new synchronization conditions for complex networks with nonlinear and nonidentical self-dynamics with switching directed communication graphs. In light of the recent works on distributed sub-gradient methods, we impose   integral convexity for the nonlinear node self-dynamics in the sense that the self-dynamics of a given node is the  gradient of some concave function corresponding to that node.  The node couplings  are assumed to be linear  but with switching directed communication graphs. Several sufficient and/or necessary conditions are established for exact or approximate synchronization over the considered complex  networks. These results show when and how nonlinear  node self-dynamics may cooperate with the linear diffusive coupling, which eventually  leads to network synchronization conditions under  relaxed connectivity requirements.
\end{abstract}


\section{Introduction}
The past few decades have  witnessed tremendous research interest in the emergence of collective behaviors  for  dynamics over complex  networks \cite{prl,vic95,mar, wubook}. The new understanding we gained lies in that certain global network-level tasks, such as  synchronization or consensus, can be achieved by local interactions  under cooperative couplings of individual node dynamics \cite{wu-chua95,wubook,jad03}. More advanced strategies have also been developed for problems like formation, swarming, optimization, and  signaling \cite{fax,mar,saber,nedic09,nedic10,jmf13}.

Synchronization problems require  the node states to asymptotically reach a common trajectory or a common value  over a network. In \cite{prl}, a master stability function method was proposed for the local synchronization of  linearly  coupled oscillators, where the dynamics of each node consists of a term of nonlinear self-dynamics and another  term of local linear couplings. In  \cite{wu-chua95}, a thorough treatment was established  for  synchronization of linear diffusive couplings. When the node self-dynamics  is nonlinear, it was shown that the coupling strength must dominant the influence of this self-dynamics in order for global synchronization \cite{physicad04,wubook}. Further extensions for linearly coupled oscillators  have been established under more restrictions on the  individual self-dynamics, e.g.,  passivity, symmetry, and linearity  \cite{nij01,nij02,li10,you11}. These works mainly  focused on fixed interaction graph and identical self-dynamics.

Efforts have also been made on synchronization under switching interactions, non-identical node self-dynamics, or nonlinear couplings, which turned out to be far more challenging    \cite{sepulchre09,bullo12}. Some recent improvements to classical synchronization results include \cite{sutac} and \cite{caoming13}. In \cite{sutac}, connectivity requirements were relaxed to jointly connected undirected graphs, where only the union of the switching communication graphs is assumed to be connected over certain intervals, for linear agent models. In \cite{caoming13}, the authors provided a graph comparison perspective, based on which some new graphical conditions were obtained for synchronization conditions with nonlinear node self-dynamics but fixed communication graphs.

The difficulty in analyzing synchronization conditions comes from the nontrivial coupling between node self-dynamics and the local interactions, as well as the coupling between different node states, especially under a switching communication graph. While without self-dynamics in each node, network synchronization falls to a  distributed consensus problem. For consensus seeking, it has been shown that various convergence conditions can be derived based on much relaxed connectivity conditions with even directed node interactions \cite{tsi,jad03,fax,ren05,caoming1,mor,lin07,shisiam}. On the other hand, it has also been shown that if the node self-dynamics can be properly designed, this node self-dynamics can cooperate with the consensus couplings leading to distributed solutions to certain network optimization problems \cite{nedic09,nedic10,elia,shitac,jmf,eben,cotes14}, which generalized the classical  incremental methods for distributed optimization  \cite{solodov,nedic01,ram}.

In this paper, we try to borrow the insights from consensus-based distributed optimization methods \cite{nedic09,nedic10,elia,shitac,cotes14}, with the aim of establishing some new synchronization conditions which can partially relax the in general strong  assumptions on the  nonlinear node self-dynamics \cite{wu-chua95,caoming13}. We assume that the network nodes have non-identical nonlinear self-dynamics as gradients of some concave functions. This allows for a new class of nonlinear self-dynamics which to the best of our knowledge has not been addressed in the literature.   The node couplings are linear diffusive   with switching directed communication graphs. Several sufficient and/or necessary conditions are established for exact or approximate synchronization of the overall node states. These results reveal when and how nonlinear  node self-dynamics may cooperate with the linear consensus coupling, which leads to synchronization conditions under much relaxed connectivity requirements to the communication graphs.

The remainder of the paper is organized as follows. In Section~2, some preliminary mathematical concepts and lemmas are introduced. In Section~3, we formulate the considered network dynamics and define the problem of interest.  Section~4 presents some results  on fixed graphs, and then  Section~5 discusses   time-varying graphs. Finally some concluding remarks are given in Section~6.

\section{Preliminaries}
In this section, we introduce some  notations and provide preliminary results that will be used in the rest of the paper.

\subsection{Directed Graphs}
A directed graph (digraph) $\mathrm
{G}=(\mathrm{V}, \mathrm{E})$ consists of a finite set
$\mathrm{V}$ of nodes and an arc set
$\mathrm{E}$, where an arc is an ordered pair of
distinct nodes of $\mathrm{V}$ \cite{god}.  An element $(i,j)\in\mathrm{E}$ describes
an arc which leaves $i$ and enters $j$.  A {\it walk} in  $\mathrm
{G}$ is an alternating sequence $\mathcal
{W}:
i_{1}e_{1}i_{2}e_{2}\dots e_{m-1}i_{m}$ of nodes $i_{\kappa}$ and
arcs $e_{\kappa}=(i_{\kappa},i_{\kappa+1})\in\mathrm{E}$ for
$\kappa=1,2,\dots,m-1$. A walk  is called a {\it path}
if the nodes of the walk are distinct, and a path from $i$ to
$j$ is denoted as $i\rightarrow j$.  A digraph $\mathrm{G}$ is called {\it undirected}
when for any two nodes $i$ and $j$, $(i,j)\in\mathrm{E}$  if and only if $(j,i)\in\mathrm{E}$; {\it strongly connected} if it contains path $i\rightarrow j$ and $j\rightarrow i$ for every pair of nodes $i$ and $j$. Ignoring the direction of the arcs, the connectivity of a undirected digraph is transformed to that of the corresponding undirected graph. A time-varying graph is defined as $\mathrm
{G}_{\sigma(t)}=(\mathrm{V},\mathrm{E}_{\sigma(t)})$ where
$\sigma:[0,+\infty)\rightarrow \mathcal {Q}$ denotes a piecewise constant function,
where $\mathcal {Q}$ is a finite set containing  all possible graphs with node set $\mathrm{V}$. Moreover, the joint graph of $\mathrm{G}_{\sigma(t)}$ in
time interval $[t_1,t_2)$ with $t_1<t_2\leq +\infty$ is denoted   as
$\mathrm{G}([t_1,t_2))= \cup_{t\in[t_1,t_2)}
\mathrm{G}(t)=(\mathrm{V},\cup_{t\in[t_1,t_2)}\mathrm
{E}_{\sigma(t)})$.

\subsection{Dini Derivatives and Limit Sets}
The upper {\it Dini
derivative} of a continuous function $h: (a,b)\to \mathds{R}$ ($-\infty\leq a<b\leq \infty$) at $t$ is defined as
$$
D^+h(t)=\limsup_{s\to 0^+} \frac{h(t+s)-h(t)}{s}.
$$
When $h$ is continuous on $(a,b)$, $h$ is
non-increasing on $(a,b)$ if and only if $ D^+h(t)\leq 0$ for any
$t\in (a,b)$. The next
result is convenient for the calculation of the Dini derivative \cite{dan,lin07}.

\begin{lemma}
\label{lemdini}  Let $V_i(t,x): \mathds{R}\times \mathds{R}^d \to \mathds{R}\;(i=1,\dots,n)$ be
$C^1$ and $V(t,x)=\max_{i=1,\dots,n}V_i(t,x)$. If $
\mathcal{I}(t)=\{i\in \{1,2,\dots,n\}\,:\,V(t,x(t))=V_i(t,x(t))\}$
is the set of indices where the maximum is reached at $t$, then
$
D^+V(t,x(t))=\max_{i\in\mathcal{ I}(t)}\dot{V}_i(t,x(t)).
$
\end{lemma}

Next, consider the following
autonomous system
\begin{equation}
\label{i1} \dot{x}=f(x),
\end{equation}
where $f:\mathds{R}^d\rightarrow \mathds{R}^d$ is a  continuous function. Let $x(t)$ be a solution of
(\ref{i1}) with initial condition $x(t_0)=x^0$. Then $\Omega_0\subset \mathds{R}^d$ is called a {\it positively invariant
set} of (\ref{i1}) if, for any $t_0\in\mathds{R}$ and any $x^0\in\Omega_0$,
we have $x(t)\in\Omega_0$, $t\geq t_0$, along  every solution $x(t)$ of (\ref{i1}).

We call $y$ a  $\omega$-limit point of $x(t)$ if there exists a  sequence $\{t_k\}$ with $\lim_{k\rightarrow \infty}t_k=\infty$ such that
$
\lim_{k\rightarrow \infty}x(t_k)=y.
$
The set of all $\omega$-limit points of $x(t)$ is called the  $\omega$-limit set of  $x(t)$,  and is denoted as $\Lambda^+\big(x(t)\big)$.
The following lemma is well-known  \cite{rou}.
\begin{lemma}\label{leminvariant}
Let  $x(t)$ be a solution of (\ref{i1}). Then  $\Lambda^+\big(x(t)\big)$ is positively  invariant. Moreover, if $x(t)$ is contained in a compact set,
then $\Lambda^+\big(x(t)\big)\neq \emptyset$.
\end{lemma}

\subsection{Convex Analysis}
A set $K\subset \mathds{R}^d$ is said to be {\it convex} if $(1-\lambda)x+\lambda
y\in K$ whenever $x\in K,y\in K$ and $0\leq\lambda \leq1$.
For any set $S\subset \mathds{R}^d$, the intersection of all convex sets
containing $S$ is called the {\it convex hull} of $S$, denoted by
${\rm co}(S)$.

Let $K$ be a closed convex subset in $\mathds{R}^d$ and denote
$|x|_K\doteq\inf_{y\in K}| x-y |$ as the distance between $x\in \mathds{R}^d$ and $K$, where $|\cdot|$
is the Euclidean norm.  There is a unique element ${P}_{K}(x)\in K$ satisfying
$|x-{P}_{K}(x)|=|x|_K$ associated to any
$x\in \mathds{R}^d$ \cite{aubin}.  The map ${P}_{K}$ is called the {\it projector} onto $K$. The following lemma holds  \cite{aubin}.
\begin{lemma}\label{lemconvex}
(i). $\langle {P}_{K}(x)-x,{P}_{K}(x)-y\rangle\leq 0,\quad \forall y\in
K$.

(ii). $|{P}_{K}(x)-{P}_{K}(y)|\leq|x-y|, x,y\in \mathds{R}^d$.

(iii) $|x|_K^2$ is continuously differentiable at  $x$ with $\nabla |x|_K^2=2\big(x-{P}_{K}(x)\big)$.
\end{lemma}

Let  $f: \mathds{R}^d\rightarrow \mathds{R}$ be a real-valued function. We call $f$ a convex function if for any $x,y\in\mathds{R}^d$ and $0\leq\lambda \leq1$,
it holds that $f\big((1-\lambda)x+\lambda y\big)\leq  (1-\lambda)f(x)+\lambda f(y)$. The following lemma states some well-known properties  for convex functions.
\begin{lemma}\label{lemfunction} Let $f:\mathds{R}^d\rightarrow \mathds{R}\in C^1$ be a convex function.

(i). $f(x)\geq f(y)+\big\langle x-y, \nabla f(y)\big\rangle$.

(ii). Any local minimum is a global minimum, i.e., $\arg \min f=\big\{z: \nabla f(z)=0 \big\}$.
\end{lemma}

The following lemma is established in \cite{shiauto} (Lemma 13, \cite{shiauto})

\begin{lemma}\label{lemmashi}
Suppose $K\subseteq \mathds{R}^m$ is a convex set and let $x_a,x_b \in \mathds{R}^m$.

(i) $\big\langle x_a-P_{K}(x_a), x_b-x_a\rangle\leq |x_a|_K \cdot \big||x_a|_K-|x_b|_K\big|$;

(ii) If $|x_a|_K > |x_b|_K$, then  $\big\langle x_a-P_{K}(x_a), x_b-x_a\rangle\leq - |x_a|_K \cdot \big(|x_a|_K-|x_b|_K\big)$.
\end{lemma}

\section{Problem Definition}

Consider a network with node set $\mathrm{V}=\{1,2,\dots,N\}$. The node interactions  are modeled by a time-varying directed graph $\mathrm{G}_{\sigma(t)}=(\mathrm{V},\mathrm
{E}_{\sigma(t)})$ with
$\sigma:[0,+\infty)\rightarrow \mathcal {Q}$ being a piecewise constant function,
where $\mathcal {Q}$ is the finite set containing  all possible digraphs over node set $\mathrm
{V}$. We assume that there is a lower bound $\tau_D>0$ between two consecutive
switching time instants of $\sigma(t)$.

A node $j$ is said to be a {\it neighbor} of $i$ at time $t$ when there is an arc $(j, i)\in \mathrm
{E}$, and we let $\mathcal{N}_i(\sigma(t))$ represent the set of agent $i$'s neighbors at time $t$.  Each node holds a state $x_i(t)\in \mathds{R}^m$. Let $a_{ij}(t)>0$ be a function marking the weight  associated with arc $(j,i)$ at time $t$. The nodes' dynamics are described as follows:
\begin{align}\label{syn}
\frac{d}{dt}{x}_i(t)=f_i\big(x_i(t)\big)+K\sum_{j\in \mathcal{N}_i(\sigma(t))}a_{ij}(t)\Big(x_j(t)-x_i(t)\Big), \ \ i=1,\dots,N,
\end{align}
 where $f_i(\cdot): \mathds{R}^m \rightarrow \mathds{R}^m$ is a continuous function denoting  the self-dynamics of node $i$ and $K\geq 0$ is a given constant. Let the weighted adjacency matrix be   denoted as $A_{\sigma (t)}$ where $[A_{\sigma (t)}]_{ij} =a_{ij}(t)$ if $j\in \mathcal{N}_i(\sigma(t))$ and  $[A_{\sigma (t)}]_{ij} =0$ otherwise.  The weighted degree  matrix is then defined as $D_{\sigma (t)}={\rm diag}(d_1(\sigma (t)),\dots,d_N(\sigma (t)))$ with $d_i(\sigma (t))=\sum_{j\in \mathcal{N}_i(\sigma(t))}a_{ij}(t)$. Then $ P_{\sigma(t)}=D_{\sigma (t)}-A_{\sigma (t)}$ is the time-varying Lapacian of the network representing the coupling of the node dynamics. For the time-varying weight function $a_{ij}(t)$, we assume that  there are $a^\ast>0$ and
$a_\ast>0$ such that $
 a_\ast\leq a_{ij}(t)\leq a^\ast,
 t\in \mathds{R}^+$.

   For the self-dynamics $f_i$, we first impose  the following assumption.



\vspace{1mm}
\noindent{\bf[{A1}]} There are $F_i: \mathds{R}^m\rightarrow \mathds{R}, i=1,\dots,N$ such that $f_i= -\nabla F_i$, where each $F_i$   is a   $C^1$ convex function with $\arg \min F_i\neq \emptyset$.
\vspace{1mm}


\begin{remark}
System (\ref{syn})  is one of the standard forms in the literature for complex network synchronization, where the first term $f_i$ represents nonlinear node self-dynamics and the second term describes linear diffusive couplings, e.g., just to name a few \cite{wu-chua95,physicad04,nij02,caoming13}. On the other hand, synchronization of networks with linear (even constant) self-dynamics but nonlinear diffusive couplings is also widely studied, e.g., the Kuramoto model \cite{kuramoto} (see \cite{bullo} for a comprehensive survey).
\end{remark}

\begin{remark}
 To investigate the synchronization of  System (\ref{syn}) under Assumption A1 is inspired by the recent developments of distributed optimization methods \cite{nedic09,nedic10}. Evidently System (\ref{syn}) is the continuous-time correspondence of the distributed   sub-gradient  algorithm proposed  in \cite{nedic09} for solving
 \begin{equation}\label{r11}
     \begin{array}{cl}
       \mathop{\rm minimize}\ & \sum_{i=1}^N F_i(z_i) \\
       \textrm{\rm subject to} & z_i\in \mathds{R}^m,  z_1=\dots=z_N.
  \end{array}
\end{equation}
Continuous-time solutions to the above problem have indeed been well studied, e.g.,  \cite{elia,elia11,eben,cotes14,shitac}, where second-order dynamics are shown to be able to derive exact solutions with fixed interaction graphs \cite{elia,cotes14}. The current paper is however more focused on  the ability of reaching a  synchronization for System (\ref{syn}), instead of the performance serving as a continuous-time solver of (\ref{r11}). In fact, clearly our results are based on   weaker assumptions, e.g., the $f_i$ are not necessarily Lipschitz and the interactions are  directed, switching, and unbalanced  (cf., \cite{nedic09,elia,elia11,cotes14}).
\end{remark}

\begin{remark}Compared to the existing  work \cite{physicad04,nij01,nij02,li10,you11,sepulchre09}: Assumption A1  does not require global Lipschitz condition, nor identical dynamics for the $f_i$. For instance, A1 allows for the case with $$
f_i(x)=-(x-m_i)^3
$$ with $m_i \in\mathds{R}$ being a constant. To the best of our knowledge, network synchronization conditions under such class of self-dynamics have not been studied in the literature.
\end{remark}

The initial time is set to be $0$. Let $x(t)=(x_1^T(t),\dots,x_N^T(t))^T\in \mathds{R}^{mN}$ be the Caratheodory solution of system (\ref{syn})  for initial condition $x^0=x(0)$. We refer to \cite{cortes} regarding the existence of the Caratheodory solution for (\ref{syn}).
 We introduce the following standard synchronization definition~\cite{caoming13}.

\begin{definition} Global synchronization   of  System (\ref{syn}) is achieved if  for all $x^0\in \mathds{R}^{mN}$, we have $\lim_{t\rightarrow +\infty}|x_i(t)-x_j(t)|=0$ for all $ i,j=1,\dots,N$.
\end{definition}

\begin{remark}
By Assumption A1 itself there might be finite-time escape for the trajectory of System (\ref{syn}), i.e., $x(t)$ approaches infinity in a finite time interval.  With suitable assumptions finite-time escape can however  be excluded.  We refer to the coming   Lemma \ref{lemmaunion}, Eq. (\ref{99}), and Lemma~\ref{lemmono}, respectively,  which guarantee the existence of $x(t)$ for the entire  $[0,\infty)$ under the corresponding conditions.
\end{remark}
\section{Fixed Undirected Graphs -- Global Results}
In this section, we consider  the possibility of  synchronization  under fixed and undirected interaction graphs. We first establish a necessary and sufficient condition for global exact synchronization, and then a global approximate synchronization condition is established.   Detailed discussions regarding the feasibility of the assumptions will be presented in the end of this section.
\subsection{Exact  Synchronization}
 We make another assumption on  the $F_i$.

\vspace{1mm}
\noindent{\bf[A2]} $\big\{z:f_i(z)=0\big\}\neq\emptyset$ is a bounded set,  and $ \big\langle x_i-P_ {\Theta_\ast}(x_i), f_i(x_i)\big\rangle \leq 0$ for all $x_i\in \mathds{R}^m$ and $i=1,\dots,N$, where $\Theta_\ast ={\rm co}\big( \bigcup_{i\in\mathrm{V}}\{z: f_i(z)=0\} \big)$.
\vspace{1mm}

Note that the inequality of Assumption A2 is not a direct consequence of convexity of the $f_i$. However, we will later  show that convexity does lead to such inequality when the argument of each $f_i$ is in $\mathds{R}$.  We present the following result.

\begin{theorem}\label{thm1}Assume that A1 and  A2 hold. Let  $\mathrm{G}_{\sigma (t)} \equiv \mathrm{G}$ for some  fixed,  undirected, and connected graph $\mathrm{G}$, and let $a_{ij}(t)\equiv a_{ji}(t)\equiv a_{ij}$ for some $a_{ij}>0,i,j=1,\dots,N$.
Then global synchronization for System (\ref{syn}) is achieved if and only if $\bigcap_{i=1}^N \big\{z:f_i(z)=0\big\}\neq \emptyset$.
\end{theorem}

{\it Proof.} (Necessity)
We first prove the necessity statement in Theorem \ref{thm1} by a contradiction argument.  Suppose
 global synchronization  is reached under the condition that  $\bigcap_{i=1}^N \big\{z:f_i(z)=0\big\}= \emptyset$. Let $x(t)$ be  a trajectory of system (\ref{syn}) $\Lambda^+(x(t))$ be its  $\omega$-limit set.

 First we show that $\Lambda^+(x(t))$ is a nonempty set.
 Introduce $$
 \theta(x(t)):= \max_{i \in \mathrm{V}} |x_i(t)|_{\Theta_\ast}^2.
 $$
  The following lemma holds (which in fact does not rely on a fixed or undirected graph), whose proof is in Appendix A.

\begin{lemma}\label{lemmaunion} Let A1 and A2 hold. Then $\theta(x(t))$ is non-increasing along each solution of System (\ref{syn}).
\end{lemma}

From the above lemma we immediately know that each trajectory $x(t)$ is contained in a compact set.   Let
\begin{align}
\mathcal{M}\doteq \big\{x=(x_1^T\dots x_N^T)^T: \ x_1=\dots=x_N;\ x_i\in \mathds{R}^m, i=1,\dots,N\big\}
\end{align}
denote the consensus manifold.  Based on Lemma \ref{leminvariant} and in view of the assumption that synchronization has been reached, we conclude  that  $\Lambda^+(x(t))\subseteq \mathcal {M}\neq \emptyset$.
Moreover, $\Lambda^+(x(t))$ is  positively invariant since (\ref{syn}) defines an autonomous system
 when the interaction  graph is fixed. This is to say, any trajectory  of system (\ref{syn})  must stay within $\Lambda^+(x(t))$ for any initial value in $\Lambda^+(x(t))$.

Now we take $y\in\Lambda^+(x(t))$. Then we have    $y=(z_\ast^T \dots z_\ast^T)^T$ for some  $z_\ast\in \mathds{R}^m$. Suppose  there exist two indices $i_1,i_2\in \{1,\dots,N\}$ with $i_1\neq i_2$ such that  $f_{i_1}(z_\ast)\neq  f_{i_2}(z_\ast)$. Consider the solution of (\ref{syn})   for initial time $0$ and initial value $y$. We have $\dot{x}_{i_1}(0)\neq \dot{x}_{i_2}(0)$. As a result, there exists a constant $\varepsilon>0$
such that $x_{i_1}(t)\neq x_{i_2}(t)$ for $t\in (0, \varepsilon)$. In other word, the trajectory will leave the consensus manifold
$\mathcal {M}$
for $(0,\varepsilon)$, and therefore will also leave the set $\Lambda^+(x(t))$. This contradicts the fact that  $\Lambda^+(x(t))$ is positively invariant.
The necessity part of Theorem \ref{thm1} has been proved.

\medskip

\noindent (Sufficiency) Note that $\mathrm{G}$ is undirected, i.e., $(i,j)\in\mathrm{E}$ if and only if $(j,i)\in \mathrm{E}$, and  $a_{ij}=a_{ji}$ for all $i$ and $j$. We  use unordered pair $\{i,j\}$  to denote the edge between node $i$ and $j$.  Denote  $\mathcal{F}(z)=\sum_{i=1}^N F_i(z)$ and  $\mathcal{F}_{\mathrm{G}}(x;K)=\sum_{i=1}^N F_i(x_i)+ \frac{K}{2}\sum_{\{j,i\}\in\mathrm{E}}a_{ij}\big|x_j-x_i\big|^2$. Denote the $N$'th Cartesian product of a set $S$ as  $S^N$. The following lemma  holds with proof given in Appendix~B.
\begin{lemma}\label{lem2}
Suppose $\bigcap_{i=1}^N \big\{z:f_i(z)=0\big\}\neq \emptyset$. Let the communication graph $\mathrm{G}$ be fixed, undirected, and  connected. Then
$\arg \min \mathcal{F}_{\mathrm{G}}(x;K)=\big( \bigcap_{i=1}^N \big\{z:f_i(z)=0\big\}\big)^N\bigcap \mathcal {M}$.
\end{lemma}

Note that
\begin{align}
K\sum\limits_{j \in
\mathcal{N}_i}a_{ij}\big(x_j-x_i\big)+ f_i\big(x_i\big)=-\nabla_{x_i} \mathcal{F}_{\mathrm{G}}(x;K).
\end{align}
As a result,
\begin{align}
\frac{d}{dt}\mathcal{F}_{\mathrm{G}}(x(t);K) = -\Big| \nabla \mathcal{F}_{\mathrm{G}}(x;K)\Big|^2
\end{align}
along each trajectory of System $(\ref{syn})$. Then  by LaSalle's invariance principle we  have
$$
\lim_{t\rightarrow \infty}{\rm dist}\big(x(t),{\arg \min \mathcal{F}_{\mathrm{G}}(x;K)}\big)=0.
$$
Lemma \ref{lem2} further  ensures
$$
\lim_{t\rightarrow \infty}{\rm dist}\bigg(x(t),\Big( \bigcap_{i=1}^N \big\{z:f_i(z)=0\big\}\Big)^N\bigcap \mathcal {M}\bigg)=0
$$
if $\mathrm{G}$ is undirected and connected.
Equivalently, global synchronization  is reached and we can even predict that each limit point of $x_i(t)$ lies in $\bigcap_{i=1}^N \big\{z:f_i(z)=0\big\}$ for all $i$.

The proof of Theorem \ref{thm1} is now complete. \hfill$\square$

\begin{remark}
We see from the proof above that the construction of $\mathcal{F}_{\mathrm{G}}(x)$    is critical because the convergence argument is based on the fact that
the gradient of $\mathcal{F}_{\mathrm{G}}(x)$ is consistent with the interaction  graph in the sense that no additional links will be introduced in the gradient.
\end{remark}

\subsection{Approximate Synchronization}
Theorem \ref{thm1} indicates that exact synchronization is impossible unless  $ \bigcap_{i=1}^N \big\{z:f_i(z)=0\big\} \neq \emptyset$ is fulfilled. In this subsection, we discuss the possibility of approximate
 synchronization   in the absence of this nonempty interaction condition.
 We introduce the  following definition.

\begin{definition}
 Global {\it $\epsilon$-synchronization}  is achieved if for all $x^0\in \mathds{R}^{mN}$, we have
\begin{equation}\label{4a}
\limsup_{t\rightarrow +\infty} \big |x_i(t)-x_j(t)\big|\leq \epsilon,\quad i,j=1,\dots,N.
\end{equation}
\end{definition}

We use the following assumption.

\vspace{1mm}

\noindent{\bf[A3]} (i) $\arg \min \mathcal{F}(z)\neq\emptyset$; (ii) $\arg \min \mathcal{F}_{\mathrm{G}}(x;K)\neq\emptyset$ for all $K\geq 0$;
(iii) $\bigcup_{K\geq 0} \arg \min \mathcal{F}_{\mathrm{G}}(x;K)$ is bounded.

\vspace{1mm}

For $\epsilon$-synchronization, we present the following result.
\begin{theorem}\label{thm2}
Assume that A1 and A3  hold. Let the interaction  graph $\mathrm{G}_{\sigma (t)} \equiv \mathrm{G}$ for some   fixed, undirected, and connected $\mathrm{G}$, and let $a_{ij}(t)\equiv a_{ji}(t)\equiv a_{ij}$ for some $a_{ij}>0,i,j=1,\dots,N$.  Then for any $\epsilon>0$,
there exists  $K_\epsilon>0$ such that global $\epsilon$-synchronization is achieved for all $K \geq K_\epsilon$.
\end{theorem}

{\it Proof.} Let's  fix $\epsilon$. Again, since
\begin{align}
K\sum\limits_{j \in
\mathcal{N}_i}a_{ij}\big(x_j-x_i\big)+ f_i\big(x_i\big)=-\nabla_{x_i} \mathcal{F}_{\mathrm{G}}(x;K),
\end{align}
the convexity of $\mathcal{F}_{\mathrm{G}}(x;K)$ ensures that
\begin{align}\label{99}
\lim_{t\rightarrow \infty}{\rm dist}\big(x(t),{\arg \min \mathcal{F}_{\mathrm{G}}(x;K)}\big)=0.
\end{align}

 Define $\tilde{\mathcal{F}}(x)=\sum_{i=1}^N F_i(x_i)$. Under Assumptions A1 and A3, we have that
\begin{align}
L_0\doteq \sup \Big\{ \big|\nabla \tilde{\mathcal{F}} (x)\big|:\ x\in \bigcup_{K\geq0} \arg \min \mathcal{F}_{\mathrm{G}}(x;K)\Big\}
\end{align}
is a finite number. We also define
\begin{align}
D_0\doteq  \sup \Big\{ \big|z_\ast-x_i\big|:\ i=1,\dots,N,\  x\in \bigcup_{K\geq 0} \arg \min \mathcal{F}_{\mathrm{G}}(x;K)\Big\},
\end{align}
where $z_\ast\in \arg \min \mathcal{F}$ is an arbitrarily chosen point.

Let $p=(p_1^T \dots p_N^T)^T \in \arg \min  \mathcal{F}_{\mathrm{G}}(x;K)$ with $p_i\in \mathds{R}^m, i=1, \dots,N$. Let $P$ be the Laplacian of the graph $\mathrm{G}$. Since the graph is undirected and connected, we can sort the eigenvalues of the matrix $P\otimes I_m$ as
$$
0=\lambda_1=\dots=\lambda_{m} <\lambda_{m+1}\leq \dots \leq\lambda_{mN}.
$$
Let $l_1\dots,l_{mN}$ be the orthonormal basis of $\mathds{R}^{mN}$ formed by the right eigenvectors of $P\otimes I_m$,
where  $l_1,\dots,l_m$ are eigenvectors corresponding to the zero eigenvalue. Suppose $p=\sum_{k=1}^{mN}c_k l_{k}$ with $c_k\in\mathds{R}, k=1,\dots,mN$.

According to (\ref{92}),  we have
\begin{align}
\Big|K(P\otimes I_m)p\Big|^2= K^2\Big|\sum_{k=m+1}^{mN} c_k \lambda_k l_k\Big|^2=K^2\sum_{k=m+1}^{mN} c_k^2 \lambda_k^2 \leq L_0^2,
\end{align}
which yields
\begin{align}\label{94}
\sum_{k=m+1}^{mN} c_k^2 \leq \Big(\frac{L_0}{K\lambda_2^\ast}\Big)^2,
\end{align}
where $\lambda_2^\ast>0$ denotes the second smallest  eigenvalue of $P$.

Now recall that
\begin{align}
\mathcal{M}\doteq \big\{x=(x_1^T\dots x_N^T)^T: \ x_1=\dots=x_N;\ x_i\in \mathds{R}^m, i=1,\dots,N\big\}.
\end{align}
is the consensus manifold. Noticing that $\mathcal{M}={\rm span} \{l_1,\dots, l_m\}$, we conclude from (\ref{94}) that
\begin{align}\label{100}
\sum_{k=m+1}^{mN} c_k^2 =\Big| \sum_{k=m+1}^{mN} c_k l_k \Big|^2=| p |_{\mathcal{M}}^2=\sum_{i=1}^N \Big|p_i-p_{\rm ave}\Big|^2\leq \Big(\frac{L_0}{K\lambda_2^\ast}\Big)^2,
\end{align}
where $p_{\rm ave}=\frac{\sum_{i=1}^N p_i}{N}$.
The last equality in (\ref{100}) is due to the fact that $\mathbf{1}_N \otimes\Big(  \frac{\sum_{i=1}^{N} p_i}{N}\Big)$ is the projection of $p$ on to $\mathcal{M}$. From (\ref{100}), $\sum_{i=1}^N \Big|p_i-p_{\rm ave}\Big|^2$ can be sufficiently small as long as $K$ is chosen sufficiently large. Noticing that $\mathcal{F}$ is a $C^1$ function, we conclude that  for any $\varsigma >0$, there is $K_1(\varsigma)>0$ such that when $K\geq K_1(\varsigma)$, there hold
\begin{align}\label{101}
\Big|p_i-p_{\rm ave}\Big|\leq \varsigma,\ i=1,\dots,N
\end{align}
and
\begin{align}
|\mathcal{F}(p_i)-\mathcal{F}(p_{\rm ave})\Big|\leq \varsigma,\ i=1,\dots,N.
\end{align}

On the other hand, with (\ref{92}), we have
\begin{align}\label{102}
\sum_{i=1}^N f_i(p_i)=\sum_{i=1}^N  f_i(p_{\rm ave}+\hat{p}_i)=0,
\end{align}
where  $\hat{p}_i=p_i-p_{\rm ave}$. Now according to (\ref{101}) and (\ref{102}),  since  $F_i\in C^1$, for any $\varsigma >0$,
there is $K_2(\varsigma)>0$ such that when $K\geq K_2(\varsigma)$,
\begin{align}
\Big|\sum_{i=1}^N f_i(p_{\rm ave}) \Big|\leq \frac{\varsigma}{D_0}.
\end{align}
This implies
\begin{align}
\mathcal{F}(p_{\rm ave})\leq \mathcal{F}(z_\ast)+|z_\ast-p_{\rm ave}|\times \Big|\sum_{i=1}^N  f_i(p_{\rm ave}) \Big|\leq \mathcal{F}(z_\ast)+\varsigma.
\end{align}

Therefore, for any $\epsilon>0$, we can take $K_0=\max\{K_1(\epsilon/2), K_2(\epsilon/2) \}$. Then when $K\geq K_0$, we have
\begin{align}
|p_i-p_j|\leq \epsilon;\ \ \mathcal{F}(p_i)\leq \min_z \mathcal{F}(z)+\epsilon
\end{align}
for all $i$ and $j$. Now with  (\ref{99}), every limit point of  system (\ref{syn})
is contained in the set $\arg \min \mathcal{F}_{\mathrm{G}}(x;K)$. Noting that $p$ is arbitrarily chosen from $\arg \min F_{\mathrm{G}}(x;K)$, $\epsilon$-synchronization  is achieved as long
as we choose $K_\epsilon\geq K_0$.  This completes the proof. \hfill$\square$

From Theorems \ref{thm1} and \ref{thm2}, we conclude that even though  without the nonempty intersection condition, it is impossible to reach exact synchronization for the considered coupled dynamics, it is still possible to find a control law  that guarantees  approximate synchronization with arbitrary accuracy.

\subsection{Assumption Feasibility}
This subsection discusses  the feasibility of Assumptions A2 and A3.

\begin{proposition}
If $\tilde{\mathcal{F}}(x)=\sum_{i=1}^N F_i(x_i)$ is coercive, i.e., $\tilde{\mathcal{F}}(x)\rightarrow \infty$ as long as $|x|\rightarrow \infty$,  then  $\big\{z:f_i(z)=0\big\}\neq\emptyset$ is a bounded set for all $i=1,\dots,N$,  and A3 holds.
\end{proposition}

{\it Proof.} First of all,  since $\tilde{\mathcal{F}}(x)=\sum_{i=1}^N F_i(x_i)$ is coercive, it follows straightforwardly that $\mathcal{F}(z)=\sum_{i=1}^N F_i(z)$ and each $F_i(z)$ are also coercive.  This implies immediately that  $\big\{z:f_i(z)=0\big\}\neq\emptyset$ is a bounded set for all $i=1,\dots,N$ and A3.(i) hold.

Next,  Observing that $\frac{K}{2}\sum_{\{j,i\}\in\mathrm{E}}a_{ij}\big|x_j-x_i\big|^2\geq 0$ for all $x=(x_1^T \dots x_N^T)^T\in \mathds{R}^{mN}$
and that $\tilde{\mathcal{F}}(x)=\sum_{i=1}^N F_i(x_i)$ is coercive,
we obtain that  $\arg \min \mathcal{F}_{\mathrm{G}}(x;K)\neq\emptyset$ for all $K\geq 0$. Thus,  A3.(ii) holds.

Finallly,  we  denote $F_\ast=\min_z \mathcal{F}(z)=\mathcal{F}(z_\ast)$.  Since $\sum_{i=1}^N F_i(x_i)$ is coercive,  there exists a constant $M(F_\ast)>0$ such that $\sum_{i=1}^N F_i(x_i)> F_\ast$ for all
$|x|>M(F_\ast)$. This implies
\begin{align}
\mathcal{F}_{\mathrm{G}}(x;K)>\mathcal{F}_{\mathrm{G}}(\mathbf{1}_N\otimes z_\ast;K)=F_\ast
\end{align}
for all $|x|> M$. That is to say, the global minimum of $\mathcal{F}_{\mathrm{G}}(x;K)$ is reached within the set $\{|x|\leq M\}$ for all $K>0$. Therefore, we have
\begin{align}
\bigcup_{K\geq 0} \arg \min \mathcal{F}_{\mathrm{G}}(x;K)\subseteq  \big\{|x|\leq M(F_\ast)\big\}.
\end{align}
This proves A3.(iii). \hfill$\square$

\begin{proposition}
Suppose $\big\{z:f_i(z)=0\big\}\neq\emptyset$ is a bounded set for all $i=1,\dots,N$ and the node state space  is $\mathds{R}$, i.e., $m=1$. Then A2 and A3  hold.
\end{proposition}

{\it Proof.}  Since each $\big\{z:f_i(z)=0\big\}$ is a finite interval when the node state is one dimensional, it is straightforward to verifying that
  $ \big\langle x_i-P_ {\Theta_\ast}(x_i), f_i(x_i)\big\rangle \leq 0$ for all $x_i\in \mathds{R}$. Thus A2 holds. We now prove A3 also holds.

\noindent (i). Let $x_i^\ast \in \arg \min F_i$.  Denote $y_\ast=\min \{x_1^\ast,\dots,x_N^\ast\}$. Then for any $i=1,\dots,N$, we have
\begin{align}
0\geq F_i(x_i^\ast)-F_i(y_\ast)\geq -(x_i^\ast -y_\ast)f_i(y_\ast)
\end{align}
according to inequality (i) of Lemma \ref{lemfunction}. This immediately yields $f_i(y_\ast)\geq  0$ for all $i=1,\dots,N$.

Thus, for any $y<y_\ast$, we have
\begin{align}
\mathcal{F}(y)-\mathcal{F}(y_\ast)\geq (y -y_\ast)\nabla \mathcal{F}(y_\ast)=-\sum_{i=1}^N (y -y_\ast) f_i(y_\ast)\geq 0,
\end{align}
which implies $F(y)\geq F(y_\ast)$ for all  $y<y_\ast$.

A symmetric analysis leads to that $\mathcal{F}(y)\geq \mathcal{F}(y^\ast)$ for all  $y>y^\ast$ with $y^\ast=\max \{x_1^\ast,\dots,x_N^\ast\}$. Therefore, we obtain
$ \mathcal{F}(y)\geq \min\{\mathcal{F}(y_\ast), \mathcal{F}(y^\ast)\}$ for all $y\neq [y_\ast, y^\ast]$. This implies that a global minimum is reached within
the interval $[y_\ast, y^\ast]={\rm co}\{x_1^\ast,\dots,x_N^\ast\}$  and A3.(i) thus follows.

\noindent (ii).  Introduce the following cube in $\mathds{R}^N$:
$$
\mathcal{C}_\ast^\eta\doteq \Big\{x=(x_1^T \dots x_N^T)^T: \ x_i \in [y_\ast-\eta, y^\ast+\eta],i=1,\dots,N\Big\},
$$
where $\eta>0$ is a given constant.

\vspace{1mm}

\noindent {\it Claim.} For any $K\geq 0$, $\mathcal{C}_\ast^\eta$ is an invariant set of System (\ref{syn}).

\vspace{1mm}

Define $\Psi(x(t))=\max_{i\in\mathrm{V}} x_i(t)$. Then based on Lemma \ref{lemdini}, we have
\begin{align}
D^+\Psi(x(t))&=\max_{i\in \mathcal{I}_0(t)} \frac{d}{dt}x_i(t)\nonumber\\
&=\max_{i\in \mathcal{I}_0(t)}  \Big[ \sum\limits_{j \in
\mathcal{N}_i}a_{ij}\big(x_j-x_i\big)+f_i\big(x_i\big) \Big] \nonumber\\
&\leq  \max_{i\in \mathcal{I}_0(t)} \Big[f_i\big(x_i\big) \Big],
\end{align}
where $\mathcal{I}_0(t)$ denotes the index set which contains all the nodes reaching the maximum for $\Psi(x(t))$.

Since
\begin{align}
0\geq F_i(x_i^\ast)-F_i(y_\ast+\eta)\geq -(x_i^\ast -y_\ast-\eta)f_i(y_\ast+\eta),\ i=1,\dots,N
\end{align}
we have $f_i( y^\ast+\eta)\leq 0$ for all $i=1,\dots,N$. As a result, we obtain
\begin{align}
D^+\Psi(x(t))\Big|_{\Psi(x(t))=y^\ast+\eta}\leq 0,
\end{align}
which implies $\Psi(x(t))\leq y^\ast+\eta$ for all $t\geq t_0$ under initial condition $\Psi(x(t_0))\leq y^\ast+\eta$. Similar analysis ensures that
$\min_{i\in\mathrm{V}} x_i(t)\geq y^\ast-\eta$ for all $t\geq t_0$ as long as $\min_{i\in\mathrm{V}} x_i(t_0)\geq y^\ast-\eta$. This proves the claim.

\vspace{2mm}

Note that every trajectory of system (\ref{syn})  asymptotically reaches $ \arg \min\mathcal{F}_{\mathrm{G}}(x;K))$. This immediately leads to
that $ \mathcal{F}_{\mathrm{G}}(x;K)$ reaches its minimum within   $\mathcal{C}_\ast^\eta$ for any $K\geq 0$ since $\mathcal{C}_\ast^\eta$ is an invariant set.
Then  A3.(ii) holds.

\noindent (iii). Since  $\arg \min F_i$ is bounded for $i=1,\dots,N$, there exist $b_i\leq d_i, i=1,\dots, N$ such that $\arg \min F_i=[b_i, d_i]$. Define $b_\ast =\min\{b_1,\dots,b_N\}$ and
$d^\ast= \max\{d_1,\dots,d_N\}$. We will prove the conclusion by showing $\arg \min F_{\mathrm{G}}(x;K) \subseteq \mathcal{C}_\ast$ for all $K\geq 0$, where
$$
\mathcal{C}_\ast\doteq \Big\{x=(x_1^T \dots x_N^T)^T: \ x_i \in [b_\ast, d^\ast],i=1,\dots,N\Big\}.
$$

Let $z=(z_1 \dots,z_N)^T\in \arg \min  \mathcal{F}_{\mathrm{G}}(x;K)$. First we show $\max\{z_1,\dots,z_N\} \leq d^\ast$ by a contradiction argument. Suppose $\max\{z_1,\dots,z_N\} > d^\ast$.

Now let $i_1,\dots, i_k$ be the nodes reaching the maximum state, i.e.,  $z_{i_1}=\dots=z_{i_k}=\max\{z_1,\dots,z_N\}$. There will be two cases.
\begin{itemize}
\item Let $k=N$. We have $z_1=\dots=z_N=y$ in this case. Then for all $i$ and $x_i^\ast \in \arg \min F_i$, we have
\begin{align}
0>F_i(x_i^\ast)-F_i(y)\geq - (x_i^\ast -y) f_i(y)
\end{align} which yields $ f_i(y)>0, i=1,\dots,N$ since $y>d^\ast$. This immediately leads to
\begin{align}
 \mathcal{F}_{\mathrm{G}}(z;K)= \mathcal{F}(y)>\min  \mathcal{F} \geq \min  \mathcal{F}_{\mathrm{G}}(z;K),
\end{align}
which contradicts the fact that $z\in \arg \min  \mathcal{F}_{\mathrm{G}}(x;K)$.
\item Let $k<N$. Then we denote  $s_\ast=\max\big \{ z_i: i\notin\{i_1,\dots,i_k\}, i=1,\dots,N \big\}$, which is actually the second largest value in $\{z_1,\dots,z_N\}$.
We define a new point $\hat{z}=(\hat{z}_1 \dots,\hat{z}_N)^T$ by $\hat{z}_i=z_i, i\notin\{i_1,\dots,i_k\}$ and
	\begin{align}
\hat{z}_i=\begin{cases}
		d^\ast, & \mbox{if $s_\ast<d^\ast$}\\
	s_\ast, & \mbox{otherwise}
	\end{cases}
\end{align}
for $i\in \{i_1,\dots,i_k\}$. Then it is easy to obtain that $ \mathcal{F}_{\mathrm{G}}(z;K)> \mathcal{F}_{\mathrm{G}}(\hat{z};K)$, which again contradicts the choice of $z$.
\end{itemize}
Therefore, we have proved that $\max\{z_1,\dots,z_N\} \leq d^\ast$. Based on a  symmetric analysis we also have $\min\{z_1,\dots,z_N\} \geq b_\ast$. Therefore, we obtain
$\arg \min \mathcal{F}_{\mathrm{G}}(x;K) \subseteq \mathcal{C}_\ast$ for all $K\geq 0$ and A3.(iii) follows. \hfill$\square$
\section{Time-varying Directed Graphs -- Global Exact and Semi-global Approximate Synchronization}
In this section,  we consider time-varying graphs. We introduce  the following definition \cite{jad03,lin07}.
\begin{definition}  $\mathrm
{G}_{\sigma(t)}$ is said to be {\it uniformly jointly strongly connected}  if there exists a constant $T>0$
such that $\mathrm{G}([t,t+T))$ is strongly connected for any $t\geq0$.
\end{definition}

We present the following result.
\begin{theorem}\label{thm3} Let  A1 hold.  Suppose   $\mathrm
{G}_{\sigma(t)}$ is uniformly jointly strongly connected and   $\bigcap_{i=1}^N \big\{z:f_i(z)=0\big\}\neq \emptyset$  contains at least one interior point.
Then global synchronization is achieved for System (\ref{syn}). In fact, for any initial value $x^0$, there exists $x_\ast\in \bigcap_{i=1}^N \big\{z:f_i(z)=0\big\}$ such that $\lim_{t\rightarrow\infty}x_i(t)=x_\ast$
for all $i\in\mathrm{V}$.
\end{theorem}

Note that the condition $\lim_{t\rightarrow\infty}x_i(t)=x_\ast$ is indeed a stronger conclusion than our definition of synchronization  as Theorem \ref{thm3} guarantees that all the node states converge to a common point. We will see from the proof of Theorem \ref{thm3} that this state convergence highly relies on the existence of an interior point
of $\bigcap_{i=1}^N \big\{z:f_i(z)=0\big\}$. In the absence of such an interior point condition, it turns out that global synchronization  still stands. We present another theorem stating the fact.

\begin{theorem}\label{thm4} Let A1 hold. Suppose   $\mathrm
{G}_{\sigma(t)}$ is uniformly jointly strongly connected and $\bigcap_{i=1}^N \big\{z:f_i(z)=0\big\}\neq \emptyset$.
Then global synchronization is achieved for System (\ref{syn}).
\end{theorem}

For $\epsilon$-synchronization under switching interactions, we present the following result.

\begin{theorem}\label{thm5} Let A1 and  A2 hold. Suppose   $\mathrm
{G}_{\sigma(t)}$ is uniformly jointly strongly connected.
Then for any $\epsilon>0$ and any initial value $x^0\in \mathds{R}^{mN}$,
there exist a sufficiently small $T_\epsilon^\dag(x^0)>0$  and a sufficiently large $K_\epsilon^\dag(x^0)$ such that  $\epsilon$-synchronization is achieved under $x^0$ for all $T\leq T_\epsilon^\dag(x^0)$ and  $K \geq K^\dag_\epsilon(x^0)$.
\end{theorem}

Note that compared to the results under discrete-time dynamics \cite{nedic09,nedic10}, Theorems \ref{thm3} and \ref{thm4} stand on quite general assumptions, which applies to the case when the $\big\{z:f_i(z)=0\big\}$ are unbounded.
Compared to Theorem \ref{thm2}, Theorem \ref{thm5} is semi-global in the sense that the control gain  $K_\epsilon^\dag(x^0)$ depends on the initial value. With switching interaction graphs, it becomes fundamentally difficult to characterize the limit set of the trajectories, and a general global result as Theorem \ref{thm2} may not  hold.

 The remaining of this section presents the proofs of the above   results. We first present some useful lemmas, and then the proofs of  Theorems \ref{thm3}, \ref{thm4}, and \ref{thm5} will be established, respectively.

\subsection{Preliminary  Lemmas}
We establish three useful lemmas in this subsection. Suppose $\bigcap_{i=1}^N \big\{z:f_i(z)=0\big\}\neq \emptyset$ and take $z_\ast\in \bigcap_{i=1}^N \big\{z:f_i(z)=0\big\}$. We define
\begin{align}
V_i(t)=\big|x_i(t)-z_\ast\big|^2,\ i=1,\dots,N,
\end{align}
and
\begin{align}
V(t)=\max_{i=1,\dots,N} V_i(t).
\end{align}
The following lemma holds, whose proof is given in Appendix C.
\begin{lemma}\label{lemmono} Let A1 hold. Suppose $\bigcap_{i=1}^N \big\{z:f_i(z)=0\big\}\neq \emptyset$.
Then along any trajectory of System  (\ref{syn}), we have $D^+V(t)\leq 0$ for all $t\in \mathds{R}^+$.
\end{lemma}

A direct consequence of Lemma \ref{lemmono} is that when $\bigcap_{i=1}^N \big\{z:f_i(z)=0\big\}\neq \emptyset$, we have
\begin{align}
\lim_{t\rightarrow \infty} V(t)=d_\ast^2
\end{align}
for some $d_\ast\geq 0$ along any trajectory of system  (\ref{syn}) with control law $\mathcal{J}_\star(n_i,g_i)$.
However, it is still unclear whether $V_i(t)$ converges or not. We establish another lemma indicating that with proper connectivity condition for the communication graph, all $V_i(t)$'s
have the same limit $d_\ast^2$. The following Lemma holds with proof given in Appendix D.

\begin{lemma}\label{lemlimit}
Let A1 hold. Suppose $\bigcap_{i=1}^N \big\{z:f_i(z)=0\big\}\neq \emptyset$ and $\mathrm
{G}_{\sigma(t)}$ is uniformly jointly strongly  connected. Then along any trajectory of System  (\ref{syn}),
we have $\lim_{t\rightarrow \infty} V_i(t)=d_\ast^2 $ for all $i$.
\end{lemma}

Finally, the next lemma shows that each $x_i(t)$ asymptotically  reaches $\arg\min F_i$  along the trajectories of system  (\ref{syn}), whose proof is in Appendix E.

\begin{lemma}\label{lemnodeoptimum}
Let A1 hold. Suppose $\bigcap_{i=1}^N \big\{z:f_i(z)=0\big\}\neq \emptyset$ and $\mathrm
{G}_{\sigma(t)}$ is uniformly jointly strongly  connected. Then along any trajectory of system  (\ref{syn}),
we have $$
\limsup_{t\rightarrow \infty} \big| x_i(t)\big|_{\arg\min F_i}=0
$$ for all $i$.
\end{lemma}

\subsection{Proofs of Statements}
\subsubsection{Proof of Theorem \ref{thm3}}
The proof of Theorem \ref{thm3} relies on the following lemma.
\begin{lemma}\label{lemunique} Let $z_1,\dots,z_{m+1}\in\mathds{R}^m$ and $d_1,\dots,d_{m+1}\in\mathds{R}^+$.
Suppose there exist solutions to equations (with variable $y$)
\begin{equation}\label{23}
	\begin{cases}
		|y-z_1|^2 =d_1;\\
		\ \ \ \ \ \  \vdots\\
|y-z_{m+1}|^2 =d_{m+1}.
	\end{cases}
\end{equation}
Then the solution  is unique if ${\rm rank}\big(z_2-z_1, \dots, z_{m+1}-z_1\big)=m$.
\end{lemma}
{\it Proof.} Take $j>1$ and let $y$ be a solution to the equations. Noticing that
$$
\langle y-z_1,y-z_1\rangle=d_1; \quad \langle y-z_j,y-z_j\rangle=d_j
$$
we obtain
\begin{align}
\langle y,z_j-z_1\rangle= \frac{1}{2}\Big(d_1-d_j+|z_j|^2-|z_1|^2\Big), \ j=2,\dots,m+1.
\end{align}
The desired conclusion follows immediately. \hfill$\square$

Let $r_\star=(r_1^T \dots r_N^T)^T$ be a limit point of a trajectory of System  (\ref{syn}).  Based on Lemma \ref{lemlimit}, we have $\lim_{t\rightarrow \infty} V_i(t)=d_\ast$ for all $z_\ast\in\bigcap_{i=1}^N \big\{z:f_i(z)=0\big\}$. This is to say,
$|r_i-z_\ast|=d_\ast$ for all $i$ and $z_\ast\in\bigcap_{i=1}^N \big\{z:f_i(z)=0\big\}$.
Since $\bigcap_{i=1}^N \big\{z:f_i(z)=0\big\}\neq \emptyset$ contains at least one interior point, it is obvious to see that we can find $z_1,\dots,z_{m+1}\in\bigcap_{i=1}^N \big\{z:f_i(z)=0\big\}$ with  ${\rm rank}\big(z_2-z_1, \dots, z_{m+1}-z_1\big)=m$ and $d_1,\dots,d_{m+1}\in\mathds{R}^+$, such that each $r_i, i=1,\dots,N$ is a solution of equations (\ref{23}). Then based on Lemma \ref{lemunique}, we conclude that $r_1=\dots=r_N$. Next, with Lemma \ref{lemnodeoptimum}, we have  $| r_i|_{\arg\min F_i}=0$. This implies that $r_1=\dots=r_N\in\bigcap_{i=1}^N \big\{z:f_i(z)=0\big\}$, i.e., global synchronization  is achieved.

We turn to state convergence. We only need to  show that $r_\star$ is unique along any trajectory of System  (\ref{syn}). Now suppose
$r_\star^1=\mathbf{1}_N\otimes r^1$ and $r_\star^2=\mathbf{1}_N\otimes r^2$ are two different limit points with $r^1\neq r^2 \in \bigcap_{i=1}^N \big\{z:f_i(z)=0\big\}$. According to the definition of a
limit point, we have that for any $\varepsilon>0$, there exists a time instant $t_\varepsilon$ such that $|x_i(t_\varepsilon)-r^1|\leq \varepsilon$ for all $i$.
Note that Lemma \ref{lemmono} indicates that the disc $B(r^1,\varepsilon)=\{y: |y-r^1|\leq \varepsilon\}$ is an invariant set for initial time $t_\varepsilon$.
While taking $\varepsilon={|r^1-r^2|}/{4}$, we see that $r^2\notin  B(r^1,|r^1-r^2|/{4})$. Thus, $r^2$ cannot be a limit point.

Now that the limit point is unique along any trajectory of System  (\ref{syn}), we denote it as  $\mathbf{1}_N\otimes x_\ast$ with $x_\ast\in \bigcap_{i=1}^N \big\{z:f_i(z)=0\big\}$.
Then we have  $\lim_{t\rightarrow\infty}x_i(t)=x_\ast$ for all $i=1,\dots,N$. This completes the proof.

\subsubsection{Proof of Theorem \ref{thm4}}
In this subsection, we prove Theorem \ref{thm4}. We need the following lemma on robust consensus, which is a special case of Theorem 4.1 and Proposition 4.10  in \cite{shisiam}.

\begin{lemma}\label{lemrobust}
Consider the following dynamics for the considered network model:
\begin{align}
\dot{x}_i=K\sum\limits_{j \in
\mathcal{N}_i(\sigma(t))}a_{ij}(t)\big(x_j-x_i\big)+w_i(t), \ i\in \mathrm{V}
\end{align}
where $K>0$ is a given constant, $a_{ij}(t)$ are weight functions satisfying our network model, and  $w_i(t)$ is a piecewise continuous  function. Let $\mathrm{G}_{\sigma(t)}$ be uniformly jointly strongly connected with respect to $T>0$.

(i). There holds   $
\lim_{t\rightarrow +\infty} \big |x_i(t)-x_j(t)\big|=0$ for all  $i,j\in \mathrm{V}$
if $\lim_{t\rightarrow \infty}w_i(t)=0, i\in \mathrm{V}$.

(ii). For any $\epsilon>0$,  there exist a sufficiently small $T_\epsilon>0$  and sufficiently large $K_\epsilon$ such that  $$
\limsup_{t\rightarrow +\infty} \big |x_i(t)-x_j(t)\big|\leq  \epsilon\|w(t)\|_{\infty}
$$
for all initial value $x^0$ when $K\geq K_\epsilon$ and $T\leq T_\epsilon$, where $ \|w(t)\|_{\infty}:= \max_{i \in\mathrm{V}} \sup_{t\in[0,\infty)} |w_i(t)|$.
\end{lemma}

Lemma \ref{lemnodeoptimum} indicates that $\limsup_{t\rightarrow \infty} \big| x_i(t)\big|_{\arg\min F_i}=0$ for all $i$, which yields
\begin{align}
\lim_{t\rightarrow \infty} f_i\big(x_i(t)\big)=0
\end{align}
for all $i$ according to Assumption A1. Then global synchronization  follows immediately from Lemma \ref{lemrobust}.(i).
Again by  Lemma \ref{lemnodeoptimum},
we further conclude that
$$
\limsup_{t\rightarrow \infty} {\rm dist}\big( x_i(t), \bigcap_{i=1}^N \big\{z:f_i(z)=0\big\}\big)=0.
$$ The desired conclusion thus follows.

\subsubsection{Proof of Theorem \ref{thm5}}

From Lemma \ref{lemmaunion} we know that  $\theta(x(t))= \max_{i \in \mathrm{V}} |x_i(t)|_{\Theta_\ast}^2$ is non-increasing under A2. As a result, we conclude that
 $$
x(t)\in \Gamma(x^0):= \Big\{z\in \mathds{R}^{mN}: \ \theta(z) \leq \theta(x^0) \Big\}
 $$
for all $t\geq 0$. Again by Assumption A2, $\Gamma(x^0)$ is a compact set. We can thus define
$$
\hbar(x^0):= \max_{i\in \mathrm{V}} \sup \Big\{ |f_i(z_i) |: \ z=(z_1 \dots z_N)^T\in \Gamma(x^0) \Big\}.
$$
Now along the trajectory $x(t)$ of (\ref{syn}) with initial value $x^0$, we have
$$
\big|f_i(x_i(t))\big|\leq \hbar(x^0)
$$
for all $t\geq 0$. Then the desired $\epsilon$-synchronization result follows immediately from  Lemma \ref{lemrobust}.(ii).
\section{Conclusions}
In light of recent works on  consensus-based distributed optimization methods, we  have established  some conditions on the synchronization problems of coupled oscillators. We assumed  that the network nodes have non-identical nonlinear self-dynamics  which are gradients of some concave functions. This allowed for functions that are not passive or globally Lipschitz.  The node interactions were  under  switching directed communication graphs. Some sufficient and/or necessary conditions are established regarding exact or approximate synchronization of the overall node states. These results revealed  when and how nonlinear  node self-dynamics can  cooperate with the linear consensus coupling and reach synchronization with much relaxed connectivity conditions. Some interesting future generalizations include  the exact convergence rate to a synchronization under strict convexity, and synchronization conditions with constrained node states.

\section*{Acknowledgment}
The authors  thank Prof. Angelia Nedi\'{c} for helpful discussions as well as  for her pointing out relevant literature.

\section*{Appendix}

\appendix
\section{Proof of Lemma \ref{lemmaunion}}

Denote $\mathcal{I}^\dag(t):=\big\{i\in\mathrm{V}:|x_i(t)|_{\Theta_\ast}^2 =\theta(x(t))\big\}$. Then from Lemma \ref{lemdini} and Lemma \ref{lemconvex}.(iii), we know
 \begin{align}\label{rr1}
 D^+\theta(x(t))=2 \max_{i\in \mathcal{I}^\dag(t)}\Big \langle x_i(t)-P_ {\Theta_\ast}(x_i(t)), f_i\big(x_i(t)\big)+K\sum_{j\in \mathcal{N}_i(\sigma(t))}a_{ij}(t)\big(x_j(t)-x_i(t)\big)\Big \rangle.
  \end{align}
Now with Lemma \ref{lemmashi}.(ii), there holds
\begin{align}\label{rr2}
\Big \langle x_i(t)-P_ {\Theta_\ast}(x_i(t)), x_j(t)-x_i(t)\Big \rangle \leq 0
  \end{align}
  for all $i\in\mathcal{I}^\dag(t) $ and $j$ due to the definition of $\mathcal{I}^\dag(t)$ and $\theta(x(t))$. Combining (\ref{rr1}), (\ref{rr2}), and Assumption A2 we further know
   \begin{align}
 D^+\theta(x(t))\leq 2 \max_{i\in \mathcal{I}^\dag(t)}\Big \langle x_i(t)-P_ {\Theta_\ast}(x_i(t)), f_i\big(x_i(t)\big)\Big \rangle \leq 0.
  \end{align}

   The desired lemma  thus follows. \hfill$\square$
\section{Proof of Lemma \ref{lem2}}
When $\bigcap_{i=1}^N \big\{z:f_i(z)=0\big\}\neq \emptyset$, it is clear that
$
\arg \min \mathcal{F} =  \bigcap_{i=1}^N \big\{z:f_i(z)=0\big\}.
$

Now take $x_\ast=(p_\ast^T \dots p_\ast^T)^T\in \big( \bigcap_{i=1}^N \big\{z:f_i(z)=0\big\}\big)^N\bigcap \mathcal {M}$, where $p_\ast\in  \bigcap_{i=1}^N \big\{z:f_i(z)=0\big\}$.
First we have $x_\ast\in \arg \min_x \sum_{i=1}^N F_i(x_i) $. Second we have $x_\ast\in \arg \min_x \frac{K}{2}\sum_{\{j,i\}\in\mathrm{E}}a_{ij}\big|x_j-x_i\big|^2$.
Therefore, we conclude that $x_\ast\in \arg \min  \mathcal{F}_{\mathrm{G}}(x;K)$. This gives
\begin{align}\label{11}
\arg \min  \mathcal{F}_{\mathrm{G}}(x;K) \supseteq\Big( \bigcap_{i=1}^N \big\{z:f_i(z)=0\big\}\Big)^N\bigcap \mathcal {M}.
\end{align}

On the other hand, convexity gives
\begin{align}\label{92}
\arg \min\mathcal{F}_{\mathrm{G}}(x;K)=\Big\{x:\ -K(P\otimes I_m) x= \Big( \big(f_1(x_1)\big)^T \dots \big(f_N(x_N)\big)^T \Big)^T\Big\},
\end{align}
where $\otimes$ represents the Kronecker product, $I_m$ is the identity matrix in $\mathds{R}^m$,  and  $P$ is the Laplacian of the graph $\mathrm{G}$. Noticing that $$
(\mathbf{1}_N^T\otimes I_m) (P\otimes I_m) =\mathbf{1}_N^TP\otimes I_m=0,
$$
where $\mathbf{1}_N=(1\dots 1)^T \in \mathds{R}^N$, we have
\begin{align}\label{8}
\Big(\mathbf{1}_N^T \otimes I_m\Big) \Big( \big(f_1(x_1)\big)^T \dots \big(f_N(x_N)\big)^T \Big)^T=\sum_{i=1}^N f_i(x_i)=0
\end{align}
for any $x\in\arg \min \mathcal{F}_{\mathrm{G}}(x;K) $.

Take $x^\ast=(q_1^T \dots q_N^T)^T\in\arg \min \mathcal{F}_{\mathrm{G}}(x;K) $. Suppose there exist two indices $i_\ast$ and $j_\ast$ such that
$$
f_{i_\ast}(q_{i_\ast})\neq f_{j_\ast}(q_{j_\ast}).
$$
Then at least one of $f_{i_\ast}(q_{i_\ast})$ and $f_{j_\ast}(q_{j_\ast})$ must be nonzero. Taking $\hat{p}\in  \bigcap_{i=1}^N \big\{z:f_i(z)=0\big\}$, we have
$$
\sum_{i=1}^N F_i(q_i)>\sum_{i=1}^NF_i(\hat{p})
$$
because for $x=(x_1^T \dots x_N^T)^T\in\arg \min  \sum_{i=1}^N F_i(x_i)$, we have $f_i(x_i)=0, i=1,\dots,N$.
Consequently, for $w_\ast=(\hat{p}^T \dots \hat{p}^T)^T$, we have
$$
\mathcal{F}_{\mathrm{G}}(x^\ast;K)>\mathcal{F}_{\mathrm{G}}(w_\ast;K)
$$
which is impossible according to the definition of $x^\ast$ so that such $i_\ast$ and $j_\ast$ cannot exist. In light of (\ref{8}), this  immediately implies
$
f_i(q_i)=0,\ i=1,\dots,N,
$
or equivalently
\begin{align}\label{10}
q_i\in \big\{z:f_i(z)=0\big\},\ i=1,\dots,N
\end{align}
 for all $x^\ast=(q_1^T\dots q_N^T)^T\in\arg \min  \mathcal{F}_{\mathrm{G}}(x)$. Therefore, we conclude from (\ref{10}) that
 $
 \sum_{i=1}^N F_i(q_i)=\sum_{i=1}^NF_i(p_\ast),
 $
 and this implies
 $$
\sum_{\{j,i\}\in\mathrm{E}}a_{ij}\big|q_j-q_i\big|^2=0
 $$
 as long as $x^\ast=(q_1^T \dots q_N^T)^T\in\arg \min  \mathcal{F}_{\mathrm{G}}(x)$. The connectivity of the communication graph thus further guarantees  that $q_1=\dots=q_N$,
so we have proved that
 $
 x^\ast\in \big( \bigcap_{i=1}^N \big\{z:f_i(z)=0\big\}\big)^N\bigcap \mathcal {M}.$ Consequently, we obtain
\begin{align}\label{12}
\arg \min  \mathcal{F}_{\mathrm{G}}(x;K) \subseteq\Big( \bigcap_{i=1}^N \big\{z:f_i(z)=0\big\}\Big)^N\bigcap \mathcal {M}.
\end{align}

The desired lemma holds from (\ref{11}) and (\ref{12}). \hfill$\square$

\section{Proof of Lemma \ref{lemmono}}
 Based on Lemma \ref{lemdini}, we have
\begin{align}\label{20}
D^+V(t)&=\max_{i\in \mathcal{I}(t)} \frac{d}{dt}V_i(t)\nonumber\\
&=\max_{i\in \mathcal{I}(t)} 2\Big\langle x_i(t)-z_\ast, \sum\limits_{j \in
\mathcal{N}_i(\sigma(t))}a_{ij}(t)\big(x_j-x_i\big)+f_i\big(x_i\big)\Big\rangle,
\end{align}
where $\mathcal{I}(t)$ denotes the index set which contains all the nodes reaching the maximum for $V(t)$.

Let $m\in\mathcal{I}(t)$. Denote
$$
Z_t=\big\{z:\ |z-z_\ast|\leq \sqrt{V(t)} \big\}
$$
as the disk centered at $z_\ast$ with radius $\sqrt{V(t)}$. Take $y=x_m(t)+(x_m(t)-z_\ast)$. Then from some simple Euclidean geometry it is obvious to see that $P_{Z_t}(y)=x_m(t)$,
where $P_{Z_t}$ is the projection operator  onto $Z_t$. Thus, for all $j\in\mathcal{N}_{m}(\sigma(t))$, we obtain
\begin{align}\label{18}
\big\langle x_m(t)-z_\ast,x_j(t)-x_m(t)\big\rangle&=\big\langle y-x_m(t),x_j(t)-x_m(t)\big\rangle\nonumber\\
&=\big\langle y-P_{Z_t}(y),x_j(t)-P_{Z_t}(y)\big\rangle\nonumber\\
&\leq 0
\end{align}
according to inequality (i) in Lemma \ref{lemconvex} since $x_j(t)\in Z_t$. On the other hand, based on inequality (i) in Lemma \ref{lemfunction}, we also have
\begin{align}\label{19}
\big\langle x_m(t)-z_\ast,f_m\big(x_m(t)\big)\big\rangle\leq F_m(z_\ast)-F_m\big(x_m(t)\big) \leq 0
\end{align}
in light of the definition of $z_\ast$.

With (\ref{20}), (\ref{18}) and (\ref{19}), we conclude that
\begin{align}
D^+ V(t) =\max_{i\in \mathcal{I}(t)} 2\big\langle x_i(t)-z_\ast, \sum\limits_{j \in
\mathcal{N}_i(\sigma(t))}a_{ij}(t)\big(x_j-x_i\big)+f_i\big(x_i\big)\big\rangle\leq  0,
\end{align}
which completes the proof. \hfill $\square$

\section{Proof of Lemma \ref{lemlimit}}
 In order to prove the desired conclusion, we just need to show $\liminf_{t\rightarrow \infty} V_i(t)=d_\ast^2$ for all $i$. With Lemma \ref{lemmono}, we conclude that $\forall\varepsilon>0, \exists M(\varepsilon)>0$, s.t.,
\begin{align}\label{24}
\sqrt{V_i(t)}\leq d_\ast+\varepsilon
\end{align}
for all $i$ and $t\geq M$.

\vspace{2mm}
{\it Claim.} For all $t\geq M$ and all $i,j\in\mathrm{V}$, we have
\begin{align}\label{25}
\big\langle x_i(t)-z_\ast,x_j(t)-x_i(t) \big\rangle\leq-V_i(t)+(d_\ast+\varepsilon)\sqrt{V_i(t)}.
\end{align}

If $x_i(t)=z_\ast$ (\ref{25}) follows trivially from (\ref{24}). Otherwise we take $y_\ast= z_\ast+ (d_\ast+\varepsilon)\frac{x_i(t)-z_\ast}{|x_i(t)-z_\ast|}$ and $B_t=\big\{z: |z-z_\ast|\leq d_\ast+\varepsilon\big \}$. Here $B_t$ is the disk centered at $z_\ast$ with radius $d_\ast+\varepsilon$, and $y_\ast$ is a point within the boundary of $B_t$ and falls the same line with $z_\ast$ and $x_{i_0}(t)$. Take also $q_\ast=y_\ast+x_i(t)-z_\ast$. Then we have
\begin{align}
\big\langle x_i(t)-z_\ast,x_j(t)-y_\ast \big\rangle=\big\langle q_\ast-y_\ast,x_j(t)-y_\ast \big\rangle=\big\langle q_\ast-P_{B_t}(q_\ast),x_j(t)-P_{B_t}(q_\ast) \big\rangle\leq 0
\end{align}
according to inequality (i) in Lemma \ref{lemconvex}, which leads to
\begin{align}
\big\langle x_i(t)-z_\ast,x_j(t)-x_i(t) \big\rangle&=\big\langle x_i(t)-z_\ast,x_j(t)-y_\ast \big\rangle+\big\langle x_i(t)-z_\ast,y_\ast-x_i(t) \big\rangle\nonumber\\
&\leq\big\langle x_i(t)-z_\ast,y_\ast-x_i(t)\big\rangle\nonumber\\
&=-V_i(t)+(d_\ast+\varepsilon)\sqrt{V_i(t)}.
\end{align}
This proves the claim.

\vspace{2mm}

Now suppose there exists $i_0\in\mathrm{V}$ with $\liminf_{t\rightarrow \infty} V_i(t)=\theta_{i_0}^2<d_\ast^2$. Then we can find a time sequence $\{t_k\}_1^\infty$ with $\lim_{k\rightarrow \infty}t_k =\infty$ such that
\begin{align}\label{28}
\sqrt{V_{i_0}(t_k)}\leq \frac{\theta_{i_0}+d_\ast}{2}.
\end{align}
We divide the rest of the proof into three steps.

\noindent {\it Step 1.}
Take $t_{k_0}>M$. We bound $V_{i_0}(t)$ in this step.

With  (\ref{25}) and inequality (i) in Lemma \ref{lemfunction}, we see that
\begin{align}\label{45}
\frac{d}{dt} V_{i_0}(t)&=2\Big \langle x_{i_0}(t)-z_\ast, \sum\limits_{j \in
\mathcal{N}_{i_0}(\sigma(t))}a_{i_0j}(t)\big(x_j-x_{i_0}\big)+ f_{i_0}\big(x_{i_0}(t)\big) \Big\rangle\nonumber\\
&\leq 2\sum\limits_{j \in
\mathcal{N}_{i_0}(\sigma(t))}a_{i_0j}(t)  \Big \langle x_{i_0}(t)-z_\ast,x_j(t)-x_{i_0}(t) \Big\rangle+F_{i_0}\big(z_\ast\big)-F_{i_0}\big(x_{i_0}(t)\big)\nonumber\\
&\leq 2(N-1)a^\ast\Big(-V_{i_0}(t)+(d_\ast+\varepsilon)\sqrt{V_{i_0}(t)}\Big),
\end{align}
for all $t\geq t_{k_0}$, which implies
\begin{align}\label{26}
\frac{d}{dt}\sqrt{V_{i_0}(t)} \leq -(N-1)a^\ast\Big(\sqrt{V_{i_0}(t)}-(d_\ast+\varepsilon)\Big),\ \ t\geq t_{k_0}.
\end{align}

In light of Gr\"{o}nwall's inequality, (\ref{28}) and (\ref{26}) yield
\begin{align}\label{30}
\sqrt{V_{i_0}(t)} &\leq e^{-(N-1)^2a^\ast T_D}\sqrt{V_{i_0}(t_{k_0})}+\Big(1-e^{-(N-1)^2a^\ast T_D}\Big)(d_\ast+\varepsilon)\nonumber\\
&\leq \frac{e^{-(N-1)^2a^\ast T_D}}{2} \theta_{i_0}+\Big(1-\frac{e^{-(N-1)^2a^\ast T_D}}{2}\Big)(d_\ast+\varepsilon)\nonumber\\
&\doteq \Lambda_\ast.
\end{align}
for all $t\in[t_{k_0}, t_{k_0}+(N-1)T_D]$ with $T_D=T+\tau_D$, where $T$ comes from the definition of uniformly jointly strongly connected graphs and $\tau_D$ represents the dwell time.

\noindent {\it Step 2.} Since the graph is uniformly jointly strongly connected, we can find an instant $\hat{t}\in[t_{k_0},t_{k_0}+T]$ and another node $i_1\in\mathrm{V}$ such that $(i_0,i_1)\in\mathrm{G}_{\sigma(t)}$ for $t\in[\hat{t}, \hat{t}+\tau_D]$. In this step, we continue to bound $V_{i_1}(t)$.

 Similar to (\ref{25}), for all $t\geq M$ and all $i,j\in\mathrm{V}$, we also have
\begin{align}\label{29}
\big\langle x_i(t)-z_\ast,x_j(t)-x_i(t) \big\rangle\leq-\sqrt{V_i(t)}\Big(\sqrt{V_i(t)}-\sqrt{V_j(t)}\Big)
\end{align}
when $V_j(t)\leq V_i(t)$. Then based on (\ref{25}), (\ref{30}), and (\ref{29}), we obtain
\begin{align}\label{31}
\frac{d}{dt} V_{i_1}(t)
&\leq 2\sum\limits_{j \in
\mathcal{N}_{i_1}(\sigma(t))}a_{i_1j}(t)  \Big \langle x_{i_1}(t)-z_\ast,x_j(t)-x_{i_1}(t) \Big\rangle\nonumber\\
&= 2\sum\limits_{j \in
\mathcal{N}_{i_1}(\sigma(t))\setminus \{i_0\}}a_{i_1j}(t)  \Big \langle x_{i_1}(t)-z_\ast,x_j(t)-x_{i_1}(t) \Big\rangle+2a_{i_1i_0}(t)  \Big \langle x_{i_1}(t)-z_\ast,x_{i_0}(t)-x_{i_1}(t) \Big\rangle\nonumber\\
&\leq 2(N-2)a^\ast\Big(-V_{i_1}(t)+(d_\ast+\varepsilon)\sqrt{V_{i_1}(t)}\Big)-2a_\ast\sqrt{V_{i_1}(t)}\Big(\sqrt{V_{i_1}(t)}-\sqrt{V_{i_0}(t)}\Big)\nonumber\\
&\leq- 2\Big((N-2)a^\ast+a_\ast\Big)V_{i_1}(t) +2\sqrt{V_{i_1}(t)} \Big((N-2)a^\ast(d_\ast+\varepsilon)+\Lambda_\ast a_\ast\Big)
\end{align}
for $t\in[\hat{t},\hat{t}+\tau_D]$, where without loss of generality we assume $V_{i_1}(t)\geq V_{i_0}(t)$ during all $t\in[\hat{t},\hat{t}+\tau_D]$.

Then (\ref{31}) gives
\begin{align}
\frac{d}{dt} \sqrt{V_{i_1}(t)}
&\leq- \Big((N-2)a^\ast+a_\ast\Big)\sqrt{V_{i_1}(t)} + \Big((N-2)a^\ast(d_\ast+\varepsilon)+\Lambda_\ast a_\ast\Big), t\in[\hat{t},\hat{t}+\tau_D]
\end{align}
which yields
\begin{align}
\sqrt{V_{i_1}(\hat{t}+\tau_D)}&\leq e^{- \big((N-2)a^\ast+a_\ast\big)\tau_D}(d_\ast+\varepsilon)+\Big(1-e^{- \big((N-2)a^\ast+a_\ast\big)\tau_D}\Big)\frac{(N-2)a^\ast(d_\ast+\varepsilon)+\Lambda_\ast a_\ast}{(N-2)a^\ast+a_\ast}\nonumber\\
&=\frac{ a_\ast\big(1-e^{- ((N-2)a^\ast+a_\ast)\tau_D}\big)}{(N-2)a^\ast+a_\ast}\times\frac{e^{-(N-1)^2a^\ast T_D}}{2} \theta_{i_0}\nonumber\\
&\ \ \ \ \ \ \ \ \ \ \ +\Big(1-\frac{ a_\ast\big(1-e^{- ((N-2)a^\ast+a_\ast)\tau_D}\big)}{(N-2)a^\ast+a_\ast}\times\frac{e^{-(N-1)^2a^\ast T_D}}{2}\Big)(d_\ast+\varepsilon)
\end{align}
again by Gr\"{o}nwall's inequality and some simple algebra.

Next, applying the estimate of node $i_0$ in Step 1 on $i_1$ during time interval $[\hat{t}+\tau_D,t_{k_0}+(N-1)T_D]$, we arrive at
\begin{align}
\sqrt{V_{i_1}(t)}&\leq \frac{ a_\ast\big(1-e^{- ((N-2)a^\ast+a_\ast)\tau_D}\big)}{(N-2)a^\ast+a_\ast}\times\frac{e^{-2(N-1)^2a^\ast T_D}}{2} \theta_{i_0}\nonumber\\
&\ \ \ \ \ \ \ \ \ \ \ +\Big(1-\frac{ a_\ast\big(1-e^{- ((N-2)a^\ast+a_\ast)\tau_D}\big)}{(N-2)a^\ast+a_\ast}\times\frac{e^{-2(N-1)^2a^\ast T_D}}{2}\Big)(d_\ast+\varepsilon)
\end{align}
for all $t\in[t_{k_0}+T_D, t_{k_0}+(N-1)T_D]$.

\noindent{\it Step 3.} Noticing that  the graph is uniformly jointly strongly connected, the analysis of steps 1 and 2 can be repeatedly applied to nodes $i_3,\dots,i_{N-1}$, and eventually we have that for all $i_0,\dots,i_{N-1}$,
\begin{align}
\sqrt{V_{i_m}\big( t_{k_0}+(N-1)T_D\big)}&\leq \Big(\frac{ a_\ast\big(1-e^{- ((N-2)a^\ast+a_\ast)\tau_D}\big)}{(N-2)a^\ast+a_\ast}\Big)^{N-2}\times\frac{e^{-(N-1)^3a^\ast T_D}}{2} \theta_{i_0}\nonumber\\
&\ \ \ \ \ \ \ \ \ \ \ +\Bigg(1-\Big(\frac{ a_\ast\big(1-e^{- ((N-2)a^\ast+a_\ast)\tau_D}\big)}{(N-2)a^\ast+a_\ast}\Big)^{N-2}\times\frac{e^{-(N-1)^3a^\ast T_D}}{2} \Bigg)(d_\ast+\varepsilon)\nonumber\\
&<d_\ast
\end{align}
for sufficiently small $\varepsilon$  because $\theta_{i_0}<d_\ast$ and
$$
\Big(\frac{ a_\ast\big(1-e^{- ((N-2)a^\ast+a_\ast)\tau_D}\big)}{(N-2)a^\ast+a_\ast}\Big)^{N-2}\times\frac{e^{-(N-1)^3a^\ast T_D}}{2} <1
$$
is a constant. This immediately leads to that
\begin{align}
V\big(t_{k_0}+(N-1)T_D\big)<d_\ast^2,
\end{align}
which contradicts the definition of $d_\ast$.

This completes the proof. \hfill$\square$

\section{Proof of Lemma \ref{lemnodeoptimum}}
With Lemma \ref{lemlimit}, we have  that $\lim_{t\rightarrow \infty} V_i(t)=d_\ast^2$ for all $i\in\mathrm{V}$. Thus, $\forall\varepsilon>0, \exists M(\varepsilon)>0$, s.t.,
\begin{align}\label{40}
d_\ast\leq \sqrt{V_i(t)}\leq d_\ast+\varepsilon
\end{align}
for all $i$ and $t\geq M$. If $d_\ast=0$, the desired conclusion follows straightforwardly. Now we suppose $d_\ast>0$.

Assume that there exists a node $i_0$ satisfying $\limsup_{t\rightarrow \infty} \big| x_{i_0}(t)\big|_{\arg\min F_{i_0}}>0$. Then we can find a  time sequence $\{t_k\}_1^\infty$ with $\lim_{k\rightarrow \infty}t_k =\infty$ and a constant $\delta$ such that
\begin{align}\label{42}
\big| x_{i_0}(t_k)\big|_{\arg\min F_{i_0}}\geq\delta, \ k=1,\dots.
\end{align}
Denote also $B_1\doteq\big\{z: |z-z_\ast|\leq d_\ast+1\big \}$ and $G_1=\max\big\{ |f_{i_0}(y)|:\ y\in B_1\big\}$.
Assumption A1 ensures that $G_1$ is a finite number since $B_1$ is  compact. By taking $\varepsilon=1$ in (\ref{40}),
we see that $x_i(t)\in B_1$ for all $i$ and $t\geq M(1)$.  As a result, we have
\begin{align}\label{41}
\Big|\frac{d}{dt}{x}_{i_0}(t)\Big|=\Big|\sum_{j\in\mathcal{N}_{i_0}(\sigma(t))} a_{i_0 j}(t)(x_j-x_{i_0})+  f_{i_0}(x_{i_0})\Big|\leq 2(n-1) a^\ast (d_\ast+1)+G_1.
\end{align}

Combining  (\ref{42}) and (\ref{41}),   we conclude  that
\begin{align}\label{43}
\big| x_{i_0}(t)\big|_{\arg\min F_{i_0}}\geq \frac{\delta}{2}, \ t\in[t_k,t_k+\tau],
\end{align}
for all $k=1,\dots$, where by definition $\tau=\frac{\delta}{2\big( 2(n-1) a^\ast (d_\ast+1)+G_1\big)}$.

Now we introduce
$$
D_\delta\doteq \min \Big\{F_{i_0}(y)-F_{i_0}(z_\ast):\  \big| y\big|_{\arg\min F_{i_0}}\geq \frac{\delta}{2}\ {\rm and}\ y\in B_1\Big\}.
$$
Then we know $D_\delta >0$ again by the continuity of $F_{i_0}$. According to (\ref{45}), (\ref{40}),  and (\ref{43}), we obtain
\begin{align}
\frac{d}{dt} V_{i_0}(t)
&\leq 2(N-1)a^\ast\Big(-V_{i_0}(t)+(d_\ast+\varepsilon)\sqrt{V_{i_0}(t)}\Big)+F_{i_0}\big(z_\ast\big)-F_{i_0}\big(x_{i_0}(t)\big)\nonumber\\
&\leq   2(N-1)a^\ast (2d_\ast+\varepsilon)\varepsilon -D_\delta,
\end{align}
for $t\in [t_k,t_k+\tau]$, $k=1,\dots$. This leads to
\begin{align}\label{46}
V_{i_0}(t_k+\tau)&\leq V_{i_0}(t_k)+\Big(2(N-1)a^\ast (2d_\ast+\varepsilon)\varepsilon -D_\delta\Big)\tau \nonumber\\
&\leq (d_\ast+\varepsilon)^2+\Big(2(N-1)a^\ast (2d_\ast+\varepsilon)\varepsilon -D_\delta\Big)\tau\nonumber\\
&< d_\ast^2
\end{align}
as long as $\varepsilon$ is chosen sufficiently small. We see that (\ref{46}) contradicts (\ref{40}). The desired conclusion thus follows. \hfill$\square$

\end{document}